\documentclass[prl,superscriptaddress,showpacs,twocolumn,aps,a4paper,amsmath,amssymb]{revtex4}
\usepackage{float}
\usepackage{dcolumn}
\usepackage{amsmath}
\usepackage{graphicx}
\usepackage{latexsym}
\usepackage{amsfonts}
\usepackage{amssymb}
\usepackage{bm}
\usepackage{color}

\DeclareGraphicsExtensions{.pdf,.gif,.jpg}

\def\a{\alpha}
\def\b{\beta}
\def\e{\epsilon}

\def\d{\delta}

\def\G{\Gamma}
\def\l{\lambda}
\def\L{\Lambda}
\def\S{\Sigma}

\def\w{\omega}

\def\bra{\langle}
\def\ket{\rangle}

\newcommand{\be}{\begin{equation}}
\newcommand{\ee}{\end{equation}}
\newcommand{\beq}{\begin{eqnarray}}
\newcommand{\eeq}{\end{eqnarray}}

\begin{document}


\title{Screening-induced negative 
differential conductance in the Franck-Condon blockade regime}

\author{E. Perfetto}

\affiliation{Dipartimento di Fisica, Universit\`a di Roma
 Tor Vergata, Via della Ricerca Scientifica 1, I-00133 Rome, Italy}
 
\author{G. Stefanucci}

\affiliation{Dipartimento di Fisica, Universit\`a di Roma
 Tor Vergata, Via della Ricerca Scientifica 1, I-00133 Rome, Italy}
\affiliation{INFN, Laboratori Nazionali di Frascati, Via E. Fermi 40,
 00044 Frascati, Italy}
\affiliation{European Theoretical Spectroscopy Facility (ETSF)}

\begin{abstract}
    
Screening effects in nanoscale junctions 
with strong electron-phonon coupling open new physical scenarios. 
We propose an accurate many-body approach to deal with the 
simultaneous occurrence of the Franck-Condon blockade and 
the screening-induced enhancement of the polaron mobility. We derive 
a transparent analytic expression for the electrical 
current: transient and steady-state features are directly interpreted and 
explained. Moreover, the interplay between phononic and electronic 
excitations gives rise to a novel mechanism of {\em negative differential 
conductance}. Experimental setup to observe this phenomenon are 
discussed.

\end{abstract}

\maketitle

{\it Introduction.---}
The excitation of quantized vibrational modes due to passage of 
electrons in a molecular junction is at
the origin of a variety of
intriguing transport phenomena~\cite{troisi}. 
In the polaronic (strong coupling) regime electrons are blocked by the 
Franck-Condon effect and tunneling occurs via excitations of coherent many-phonon 
states~\cite{fcb}. 
This remarkable charge-transfer process engenders vibrational sidebands in the differential 
conductance $dI/dV$, as recently observed in state-of-the-art 
experiments on carbon nanotube quantum dots (QD)~\cite{vonoppen}.
A proper treatment of Coulomb charging and nuclear 
trapping already explains several features of the measured $dI/dV$.
Nevertheless, low-dimensional leads screen a charged QD by 
accumulating holes in a considerably extended portion nearby the 
contacts, thus enhancing the
electrical current to a large extent  (Coulomb 
deblocking)~\cite{mahan,borda,goldstein,irlm}. 
A quantitave assessment of screening effects in polaronic transport 
is therefore necessary
before an exhaustive interpretation of the experimental outcomes can 
be given.

This Letter contains methodological and conceptual advances on 
the transport properties of screened polarons. 
We put forward an accurate and still simple method 
to calculate the relaxation dynamics as well as the
steady-state characteristics of biased and/or gated
QDs. The key quantity is the polaron decay 
rate for which we derive a transparent analytic expression, 
highlighting the impact of the electron-electron (ee) interaction on 
systems with  electron-phonon (ep) coupling.
So far numerical simulations have been limited to ep 
interacting systems and, for all available data, we find 
excellent agreement~\cite{rabani,albrecht,wilner,thossexact}. 
In particular the extraordinary long-transient dynamics recently 
discovered in Ref.~\cite{albrecht}  
is faithfully reproduced. The simultaneous presence of ee and ep 
interactions opens new scenarios.
Relaxation still occurs through a 
long-lasting sequence of blocking-deblocking
events but the distinctive spikes in the transient current become 
much more pronounced. Noteworthily, the Coulomb deblocking has unexpected repercussions on the 
steady-state. Besides a substantial raising of the phonon-assisted 
current steps, regions of {\em Negative Differential Conductance} (NDC) 
are found in the $dI/dV$.
The NDC  is neither related to the 
asymmetry of the junction~\cite{sassetti,thoss}, nor to the finite bandwith of the 
leads~\cite{nitzan} or range of the tunneling 
amplitude~\cite{zazunov}, and disappears if the ep and ee interactions are 
considered
separately. This novel mechanism, which is of interest on its own,
complements the current understanding~\cite{sassetti} of NDC observed  in QDs~\cite{vonoppen}.

{\it Model.---}
We consider a single-level QD symmetrically connected to two 
semi-infinite 
one-dimensional leads
of length $\mathcal{L}$. Electrons on the QD
are coupled to a vibrational mode and, 
at the same time, to electrons in the leads. The Hamiltonian 
(in standard notation) reads
\beq
\hat{H}&=&t_{w}\sum_{\a,x=0}^{\infty}(\hat{c}^{\dagger}_{\a x}\hat{c}_{\a 
x+1}+\mathrm{h.c.})+ T_{l} \sum_{\a}( \hat{c}^{\dag}_{\a 0}\hat{d} +\mathrm{h.c.} ) 
\nonumber \\
&+& \e_{d} \hat{n}_{d}+ \omega_{0}\hat{a}^{\dagger}\hat{a}  + 
\lambda \hat{n}_{d} (\hat{a}^{\dag}+\hat{a})  +
U \hat{n}_{d}\sum_{\a}\hat{n}_{\a 0}, \; \; \; \;
\label{eq1}
\eeq
where $\a=L,R$ labels the left and right lead, 
$\hat{n}_{d}=\hat{d}^{\dag}\hat{d}$ and $\hat{n}_{\a 0}=
\hat{c}^{\dagger}_{\a 0}\hat{c}_{\a 0}$.
The system is driven out of equilibrium by the sudden switch-on 
of an external bias
$\hat{H}_{V}= \sum_{\a} V_{\a} \hat{N}_{\a}$, with
$\hat{N}_{\a}=\sum_{x} \hat{n}_{\a x}$ and $V=V_{L}-V_{R}$ the  
voltage drop.

At half-filling and for $V$ much smaller than the bandwidth
we can make the wide band limit approximation and consider the 
continuum version of $\hat{H}$ with a frequency independent 
tunneling rate $\G=2T_{l}^{2}/t_{w}$. 
Electrons close to the 
Fermi energy have linear dispersion $\e_{k}=v_{F}k$, with $v_{F}=
2t_{w} a$  the Fermi velocity and $a$ the lattice spacing. 
Since $k$ 
can be either positive or negative the first term of Eq. (\ref{eq1}) takes the Dirac-like 
form~\cite{boulat} $-\sum_{\a} i\a v_{F} \int dx \, \hat{\psi}^{\dagger}_{\a}(x) 
\partial_{x}\hat{\psi}_{\a}(x)$,
where $\hat{\psi}_{\a}(x)$ destroys
an electron in position $x$ of lead $\a$. 
In a similar way one can work out the other terms. The continuum 
model is obtained by
replacing $\hat{c}_{\a x} \to  \hat{\psi}_{\a}(x)$, $\sum_{x} \to \int dx$,
and by rescaling the model parameters according to $U \to u\equiv  
aU$ and $T_{l} \to t_{l}\equiv T_{l}\sqrt{2v_{F}/t_{w}}$.
We then bosonize the field operators as \cite{giamarchi}
$
\hat{\psi}_{\a}(x)=\eta_{\a} F
e^{-2\sqrt{\pi}\,i\a\hat{\phi}_{\a}(x)},
\label{bospsi}
$
with $\eta_{\a}$ the anticommuting Klein factor, $F=(\L/2\pi 
v_{F})^{1/2}$ ($\L$ is a high-energy cutoff~\cite{nota}) and  boson field 
\be
\hat{\phi}_{\a}(x)= i \a \sum_{q>0}\zeta_{q}
(\hat{b}^{\dagger}_{\a q }e^{-i\a qx}-\mathrm{h.c.})
- \sqrt{\pi}x\hat{N}_{\a}/ \mathcal{L}.
\label{bosf}
\ee
In Eq. (\ref{bosf}) the quantity $\zeta_{q}=e^{- 
\frac{v_{F}q}{2\L}}/\sqrt{2\mathcal{L}q}$. Pursuant to the 
bosonization the lead density reads
$\hat{n}_{\a}(x)=-\partial_{x}\hat{\phi}_{\a}(x)/\sqrt{\pi}$,
and the continuum Hamiltonian becomes (up to a renormalization 
of $\e_{d}$ that vanishes when $\mathcal{L}\to\infty$~\cite{irlm})
\beq
\hat{H}&=&\sum_{\a , q>0} v_{F} q \hat{b}^{\dagger}_{\a q} 
\hat{b}_{\a q} +\epsilon_{d}\hat{n}_{d} 
+ \omega_{0}\hat{a}^{\dagger}\hat{a}
\nonumber\\
&+&
t_{l} \sum_{\a}
\left[
\frac{\eta^{\dagger}_{\a}}{\sqrt{2\pi} }
e^{-2\sqrt{\pi}\sum_{q>0}\zeta_{q}
(\hat{b}^{\dagger}_{\a q}- \hat{b}_{\a q})} \hat{d} +  \mathrm{h.c.}
\right] \nonumber \\
&+&  \hat{n}_{d} \left[ 
\lambda  (\hat{a}^{\dag}+\hat{a}) -u\sum_{\a ,q>0}
\frac{\zeta_{q}q}{\sqrt{\pi}}\,
(\hat{b}^{\dagger}_{\a q} +\hat{b}_{\a q}) \right].
\label{hboson}
\eeq
Next we perform a Lang-Firsov transformation 
$\hat{H}'=\hat{\mathcal{U}}^{\dagger}\hat{H} \hat{\mathcal{U}}$
to eliminate the ep and ee coupling (third line of Eq. (\ref{hboson})).
This is achieved by the unitary operator (from now on sums are over $q>0$)
\be
\hat{\mathcal{U}}=\mathrm{exp}[-\frac{\l}{\w_{0}}
(\hat{a}^{\dag}-\hat{a})+2\sqrt{\pi}u\sum_{\a q} 
\frac{\zeta_{q}}{2\pi v}
(\hat{b}^{\dagger}_{\a q }-\hat{b}_{\a q}  ) ]\hat{n}_{d}.
\ee
In the explicit form of the transformed Hamiltonian
\be
\hat{H}'=\sum_{\a q} v_{F} q \hat{b}^{\dagger}_{\a q}
\hat{b}_{\a q} + \omega_{0}\hat{a}^{\dagger}\hat{a}
+\tilde{\epsilon}_{d}\hat{n}_{d}  
+
t_{l}\sum_{\a}
\left[ \hat{f}^{\dag}_{\a 0} \hat{d} +  \mathrm{h.c.} 
\right] 
\label{hambos}
\ee
 the screened polaron field 
\be
\hat{f}_{\a x}=\eta_{\a} F
e^{-\frac{\l}{\w_{0}}(\hat{a}^{\dag}-\hat{a})+ 2\sqrt{\pi}\sum_{\b 
q}\zeta_{q}  W_{\a \b }
(\hat{b}^{\dagger}_{\b q} e^{-i\a qx}- \hat{b}_{\b q} e^{i\a qx})}
\ee
evaluated in $x=0$ appears. In these equations $\tilde{\e}_{d}=
\e_{d}-\frac{\l^{2}}{\w_{0}}-u^{2}\sum_{q}
\frac{e^{-v_{F}q/\L}}{\pi v_{F} \mathcal{L}}$, $W_{RR}=W_{LL}=1-u/(2\pi v_{F} )$ and 
$W_{RL}=W_{LR}=-u/(2\pi v_{F} )$.
For $t_{l}=0$ we have two eigenstates with zero bosons, $|n_{d}=0,1\ket$, corresponding to 
QD occupation $n_{d}$. For $\tilde{\e}_{d}< 0$ ($\tilde{\e}_{d}> 
0$) the one 
with $n_{d}=1$ ($n_{d}=0$) is the ground state. In the following we 
consider  the system initially uncontacted ($t_{l}=0$) 
and then switch on contacts and bias.

{\it Equations of motion.---}
The advantage of working with $\hat{H}'$ is that ep and ee 
correlations are included through a calculable, transparent 
self-energy.
We define the QD Green's function on 
the Keldysh contour~\cite{svlbook} as
$G(z,z')=\frac{1}{i} \bra {\cal T} {\hat{d}(z) \hat{d}^{\dag}(z')} 
\ket$, 
where ${\cal T}$ is the contour ordering and operators are in the 
Heisenberg picture with respect to 
$\hat{H}+\hat{H}_{V}$ ($\hat{H}_{V}$ does not change after the 
Lang-Firsov transformation); the average is taken over $|n_{d}\ket$.
The QD Green's function satisfies the equation of motion 
\be
(i\partial_{z} - \tilde{\e}_{d}) G(z,z')
= \d(z,z')+ t_{l}\sum_{\a} G_{\a 0}(z,z'),
\label{eomg}
\ee 
where 
$G_{\a x} (z,z')=\frac{1}{i}\bra {\cal T}{\hat{f}_{\a x}(z)\hat{d}^{\dag}(z') }\ket$ is the 
QD-lead Green's function which in turn satisfies 
\beq
&&\left(i\partial_{z}+i\a v_{F} \partial_{x}-
i\omega_{0}\l\partial_{\l}-V_{\a}\right) G_{\a x} (z,z') 
\nonumber \\
&&=t_{l} \sum_{\b}\frac{1}{i}\bra {\cal T} \left[   
\hat{f}_{\b 0 }^{\dag} \hat{d}
+ \mathrm{h.c.} 
, \hat{f}_{\a x } 
\right](z) \hat{d}^{\dag} (z') \ket .\quad \, 
\label{eom2}
\eeq

The central approximation of our truncation scheme consists in 
replacing the 
average on the  r.h.s. of Eq. (\ref{eom2}) with 
$\bra \left( \hat{f}_{\a 0 }^{\dag}  \hat{f}_{\a x }+\hat{f}_{\a x} 
\hat{f}_{\a 0 }^{\dag}  \right)(z) \ket_{0} G(z,z') $
where $\bra \ldots \ket_{0}$ 
signifies that operators are in the Heisenberg picture with respect 
to the uncontacted but biased Hamiltonian.
This approximation corresponds to discard virtual tunneling processes between two consecutive 
ep or ee scatterings and, therefore, becomes exact for $t_{l}=0$. 
Unlike other truncation schemes~\cite{nota4}, however, 
also the noninteracting case ($\l=U=0$) is {\em exactly} recovered.

We define $ g_{\a x \a x'}(z,z')=\frac{1}{i} \bra {\cal T}{\hat{f}_{\a x }(z)  
\hat{f}_{\a x' }^{\dag}(z') } \ket_{0}$ and solve the equation of 
motion for $G_{\a 
x}$. Inserting this $G_{\a x}$ into Eq. (\ref{eomg}) yields
\be
(i\partial_{z}-\tilde{\e}_{d})G(z,z') - \int d\bar{z} \sum_{\a} 
\Sigma_{\a}(z,\bar{z}) G(\bar{z},z')=\d(z,z') ,
\label{eom4}
\ee
where $\Sigma_{\a}(z,z') =t_{l}^{2}g_{\a 0 \a 0}(z,z') $
is a correlated embedding self-energy whose greater/lesser components 
are related to the decay rate for an added/removed polaron.
In fact, $\Sigma^{>}_{\a}(t,t')$ is proportional to the 
amplitude for an electron in the QD to tunnel in lead $\a$ at 
time $t'$, explore virtually the lead for a time $t-t'$, and tunnel 
back to the QD at time $t$. A similar intepretation applies to  
$\Sigma^{<}_{\a}$.
Using the Langreth rules~\cite{svlbook} we convert
Eq. (\ref{eom4}) into a coupled system of Kadanoff-Baym
equations~\cite{dvl.2007,mssvl.2009} which can be solved
numerically once an expression for $\Sigma_{\a}$ is given. Remarkably 
the greater/lesser components of $\Sigma_{\a}$
have a simple analytic form
\beq
\Sigma^{\lessgtr}_{\a}(t-t')=\pm 
\frac{i\L\G e^{-g}}{4\pi } 
\frac{ e^{ge^{\pm i\w_{0}(t-t')}}}{[1 \mp \L(t-t')]^{\b}}
e^{-iV_{\a}(t-t') },
\label{sigmalutt}
\eeq
with ratio $g=(\l/\w_{0})^{2}$ and $u$-dependent 
exponent  $\b=1+\frac{u(u-2\pi 
v_{F})}{2\pi^{2}v_{F}^{2}}$. Equation (\ref{sigmalutt}) is our first 
main result.

\begin{figure}[tbp]
\includegraphics[width=9.25cm]{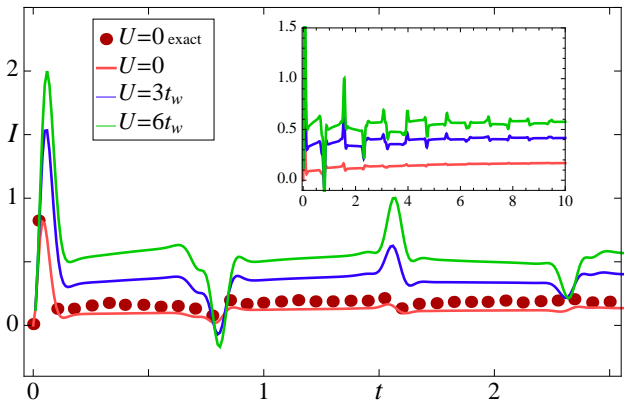}
\caption{TD current for different $U$ at 
fixed  $\l$, and initial QD occupancy $n_{d}=1$.
For $U=0$, exact data from Ref. 
\cite{albrecht} are also displayed (circles). The parameters are $\l=16$, $\w_{0}=8$, $V=26$, 
$\tilde{\e}_{d}=-10$, $\L=100$. Units: $10^{-1}\G/2$ for energies 
and $(\G/2)^{-1}$ for times.  
Inset: $I(t)$ for long propagation times (no within reach 
of current numerical techniques). 
}
\label{fig1}
\end{figure}

\begin{figure}[tbp]
\includegraphics[width=8.3cm]{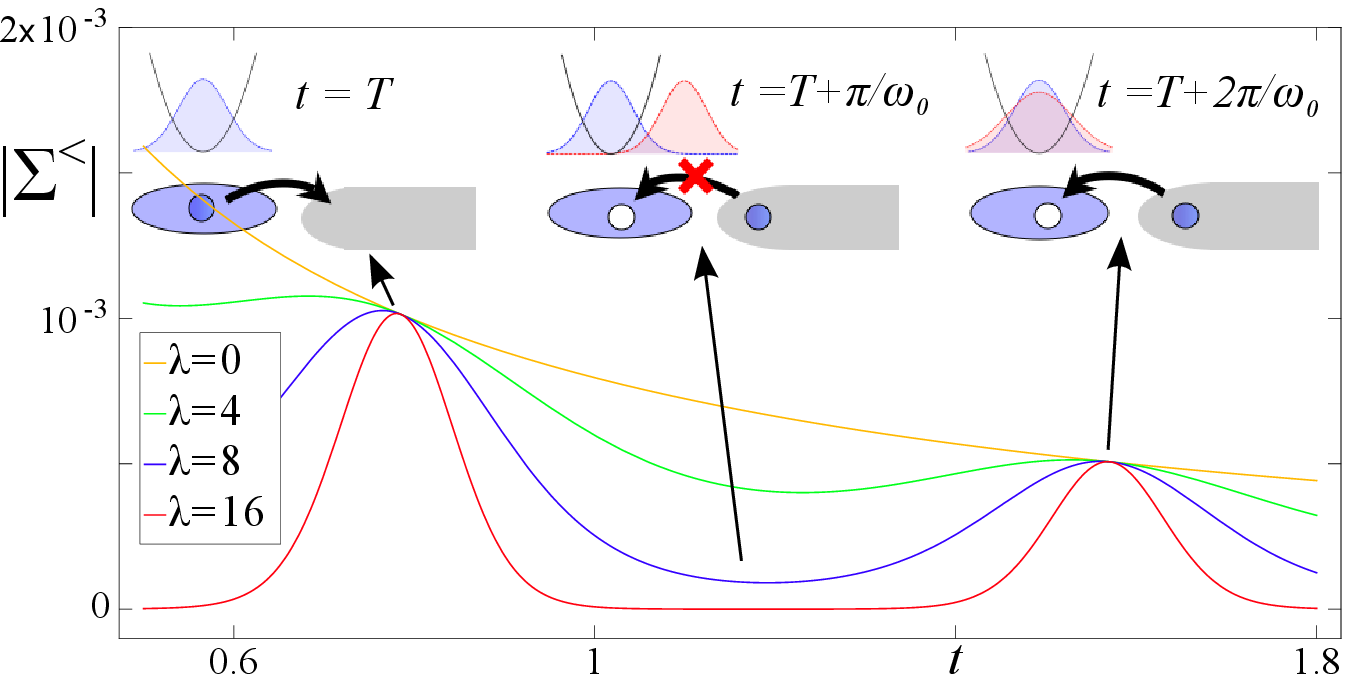}
\includegraphics[width=8.3cm]{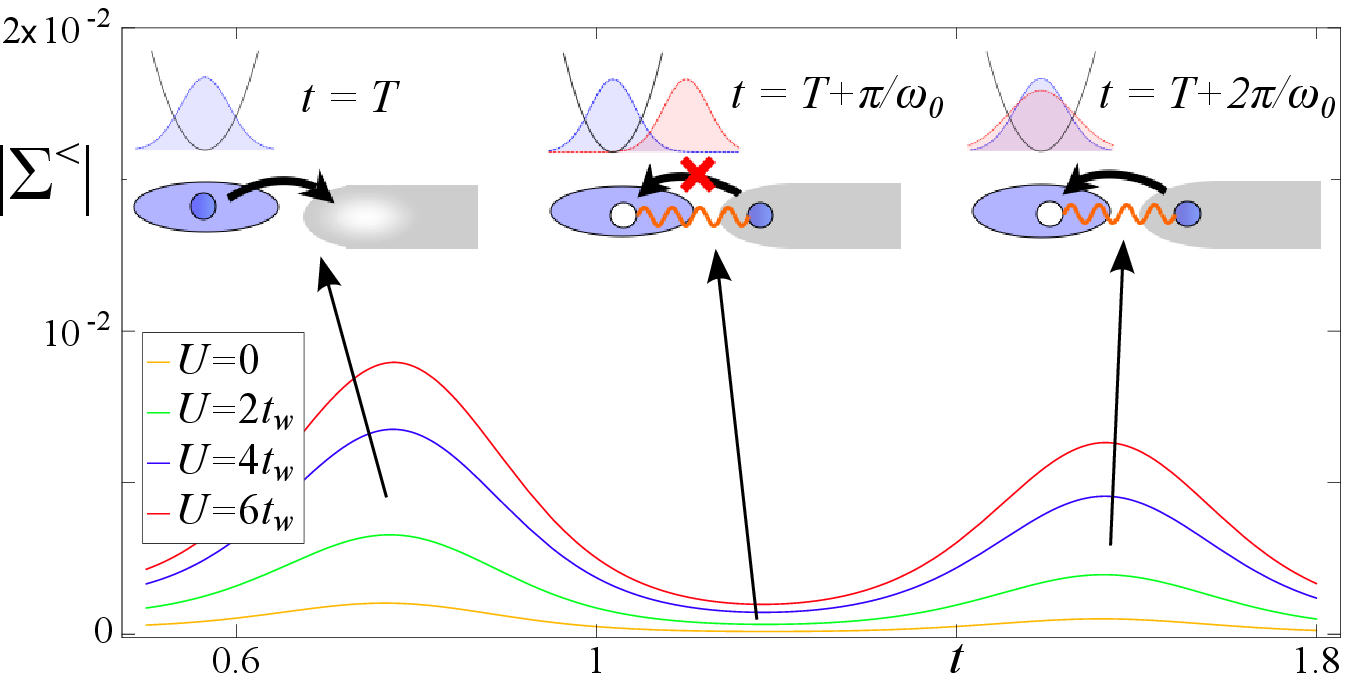}
\caption{Modulus of the TD self-energy $|\S^{<}_{\a}(t)|$
for different $\l$  at $U=0$ (top panel), and for 
different $U$ at $\l=8$ (bottom panel). 
The parameters are $\w_{0}=8$, $\tilde{\e}_{d}=0$ and
$\L=100$. Units: $\L \G $ for $|\Sigma^{<}|$,
 $\G$ for energies and $\G^{-1}$ for times.  }
\label{fig2}
\end{figure}

{\it Transient regime.---}
From the solution of Eq. (\ref{eom4}) we can extract the 
time-dependent (TD) QD density 
as well as the TD current $I_{\a}(t)$ at the $\a$ interface
\be
I_{\a}(z)=\int d\bar{z} \,\Sigma_{\a}(z,\bar{z}) G(\bar{z},z) 
+\mathrm{h.c.}\quad
\label{tdcurr}
\ee
We apply a symmetric bias $V_{L}=-V_{R}=V/2$ and calculate 
$I(t)=[I_{L}(t)+I_{R}(t)]/2$ for the parameters of Fig. \ref{fig1}.
As anticipated the $U=0$ curve is almost on top of the diagrammatic 
Monte Carlo simulation~\cite{albrecht}. 
The TD current displays quasi-stationary plateaus between 
two consecutive times $2n \pi / \w_{0}$; around these times
we see sharp spikes. For $U>0$  we observe a 
significant enhancement of the current; the plateaus bend and the 
amplitude of the spikes increases. We understand this 
peculiar transient behavior by inspecting the self-energy 
in Eq. (\ref{sigmalutt}).
In the top panel of Fig. (\ref{fig2}) we plot $|\Sigma^{<}(t)|$ for increasing 
$\l$ at $U=0$. The effect of the ep interaction is twofold:  
an overall suppression 
proportional to $e^{-g}$ and a  
modulation of period $2\pi/\w_{0}$ (coming from the double exponential 
$e^{e^{i\w_{0}t}}$). Physically (see 
cartoon in the top panel of Fig. \ref{fig2}), if we start at time 
$t=T$ with one 
electron on the QD the phonon cloud is centered around the 
minumum at $x \simeq \l/\w_{0}^{2}$ of the harmonic potential. 
The large $|\S^{<}(T)|$ favors the transfer of the electron from the 
QD to the leads
causing a sudden shift of the minumum to $x= 0$. 
At this point the polaron (electron+cloud) cannot hop back to 
the QD  since the overlap between the shifted phonon-cloud 
wavefunctions is negligible (small $|\S^{<}(t)|$). Only after a dwelling 
time of order $2\pi / \w_{0}$ this overlap is again sizable, the electron returns to the 
QD (large $|\S^{<}(T+2\pi / \w_{0})|$) and the cycle restarts. The physical 
interpretation offered by Eq. 
(\ref{sigmalutt}) enables us to 
explain the structure of the transient, how the system approaches the 
Franck-Condon blockade (FCB) regime and how screening effects change the 
picture.
\begin{figure}[b]
\includegraphics[width=4.25cm]{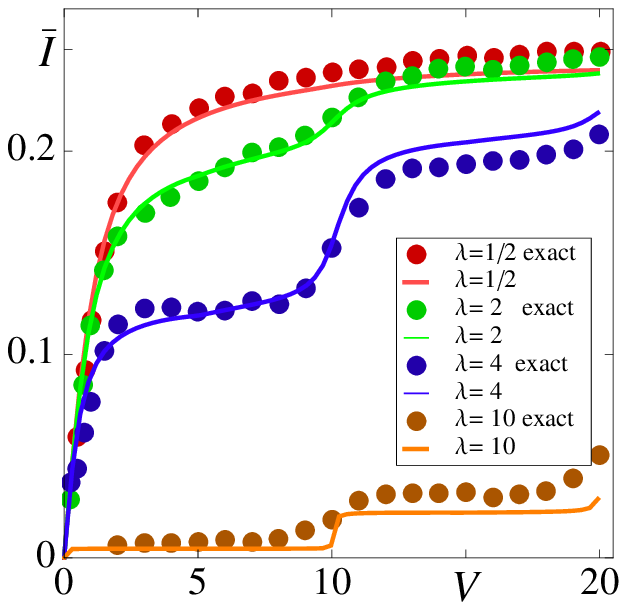}
\includegraphics[width=4.25cm]{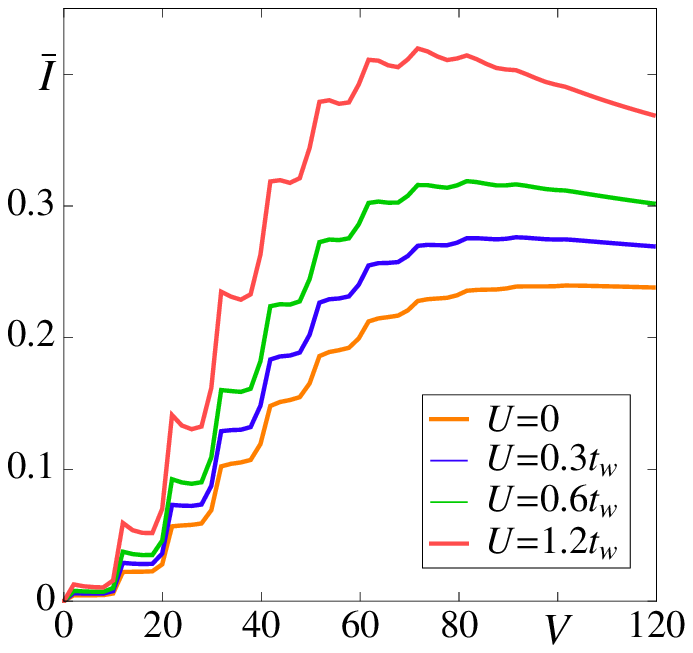}
\caption{{\it I-V} curve for different $\l$  at 
$U=0$ (left panel),  and for 
different $U$ at $\l=10$ (right panel). 
The parameters are $\w_{0}=5$, $\tilde{\e}_{d}=0$ and
$\L=1000$.
For $U=0$ (left panel), exact data from Ref. 
\cite{rabani} are also displayed (circles).
All energies in units of $\G$.
}
\label{fig3}
\end{figure}
\begin{figure*}[t]
\centering
\includegraphics*[width=.3\textwidth]{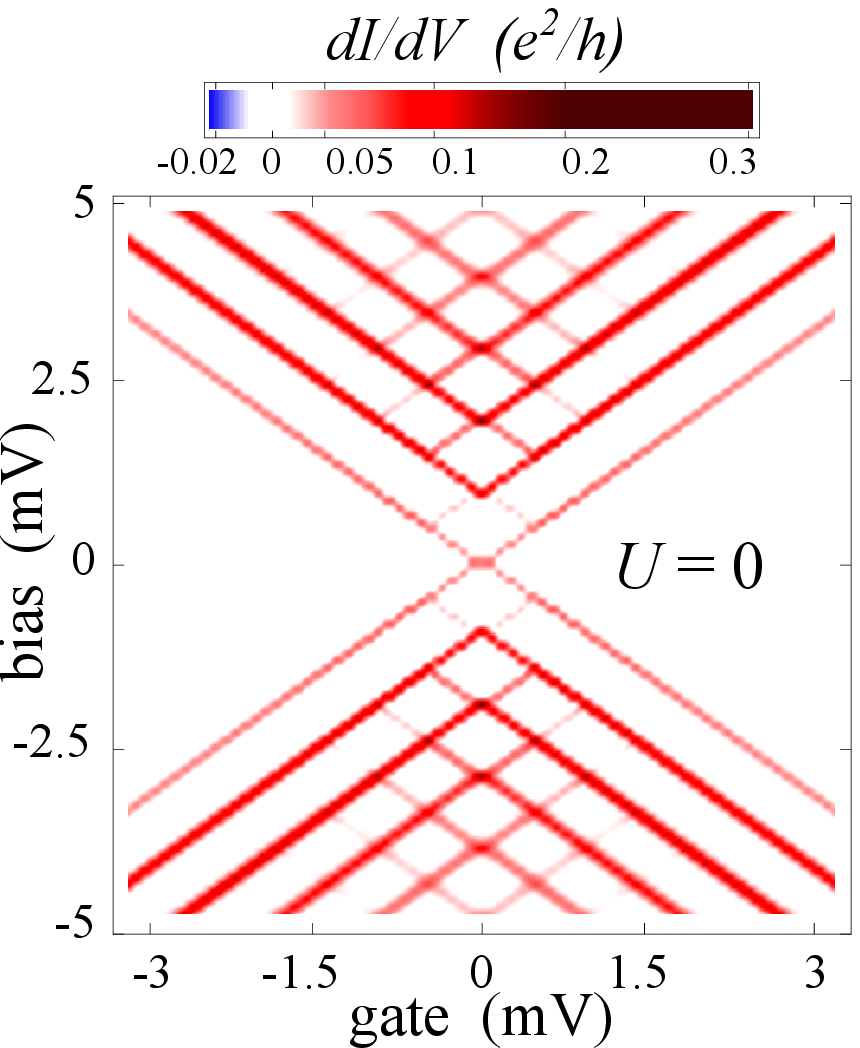}
\includegraphics*[width=.3\textwidth]{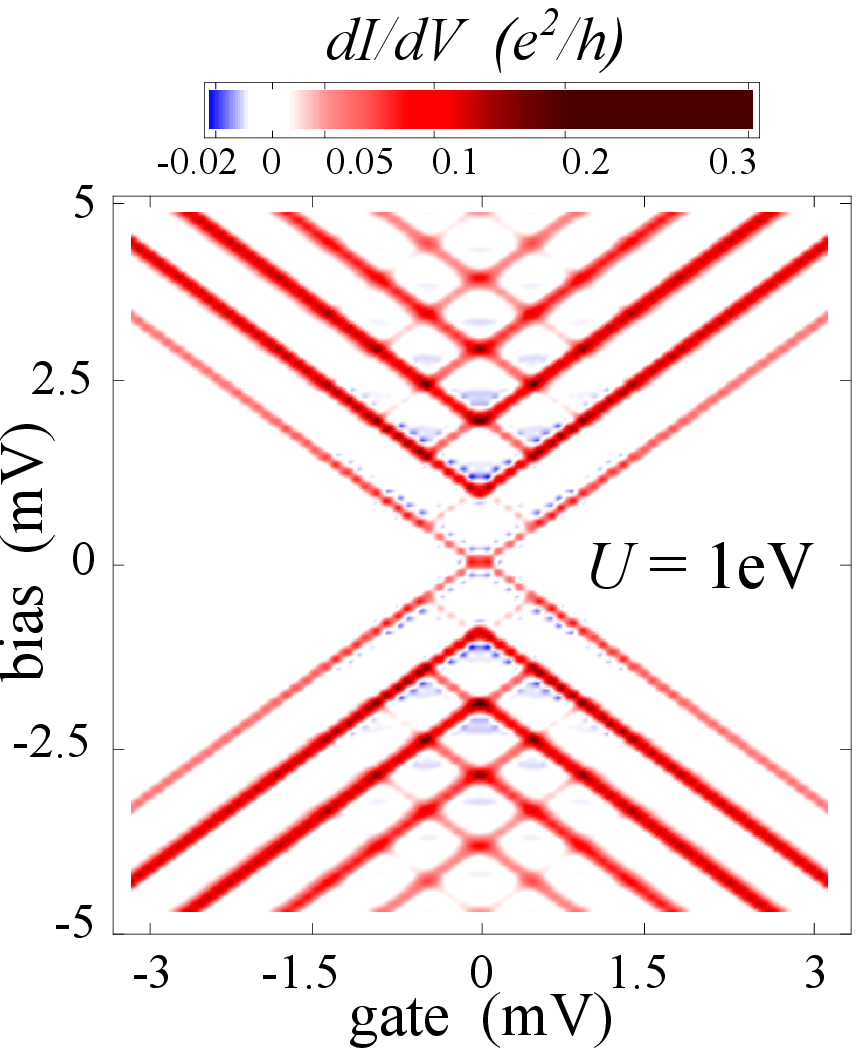}
\includegraphics*[width=.3\textwidth]{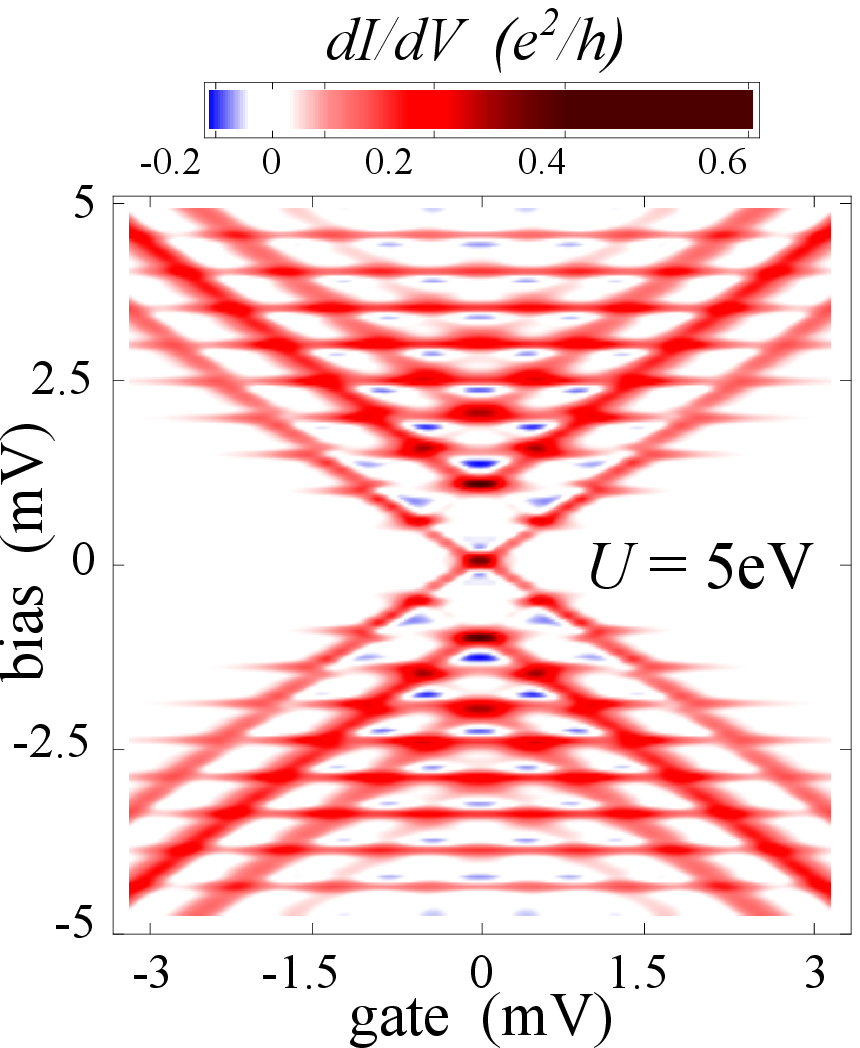}
\caption{Contour plot of the differential conductance $d I/dV$ 
as a function of the gate $\tilde{\e}_{d}$ and bias $V$, for three 
different the dot-lead repulsion: $U=0$ (left panel), $U=1$ 
eV (middle panel), $U=5$ eV (right panel). The rest of the parameters 
are specified in the text.}
\label{fig4}
\end{figure*}
Indeed, a nonvanishing $U$   modifies
the envelope of $|\S^{<}|$ from the noninteracting power-law $ 1/t$ to 
$ 1/t^{\b}$, see bottom panel of Fig. \ref{fig2}. 
According to the cartoon 
an electron in the QD causes a depletion of charge in the vicinity 
of the interface, thus facilitating the tunneling~\cite{mahan,borda,goldstein,irlm}.
Similarly, when the electron is in the leads the hole left on the QD 
acts as an attractive potential and the probability to tunnel back 
increases. This explains the enhancement of $I(t)$ in Fig. \ref{fig1}.

{\it Steady-state.---}
In the steady-state regime $G$ and $\S_{\a}$ depend only on the time 
difference and can be Fourier transformed. The 
steady current $\bar{I}$ is given by 
a Meir-Wingreen-like formula~\cite{irlm,wingreen}
\beq
\bar{I}=
\int
\frac{d\omega}{2\pi} \frac{\Sigma^{>}_{L}(\omega) \Sigma^{<}_{R}(\omega) -
\Sigma^{<}_{L}(\omega) \Sigma^{>}_{R}(\omega) }
{|\omega - \tilde{\e}_{d} -\sum_{\a}\Sigma_{\a}^{\rm 
R}(\omega)|^{2}},
\label{meir}
\eeq
where the explicit expression for the self-energy in frequency space is
\be
\Sigma^{\lessgtr}_{\a}(\w)=\pm i\frac{ \G e^{-g}}{2\G(\b)}
\sum_{n=0}^{\infty}\frac{g^{n}}{n!}
|\w^{\lessgtr}_{\a n} / \L|^{\b-1} \theta(\pm \w^{\lessgtr}_{\a n}),
\label{selfw}
\ee
with $\w^{\lessgtr}_{\a n}=\w \pm n \w_{0}-V_{\a}$, $\G(\b)$ the 
Euler-Gamma function and $\theta$ the Heavyside step function~\cite{nota2}. 

In Fig. \ref{fig3} we show the {\it I-V} curve for different $\l$  at 
$U=0$ (left panel), and for 
different $U$ at fixed $\l=10$ (right panel). 
The former is benchmarked against real-time path-integral Monte Carlo 
results~\cite{rabani}. Again we find good quantitative agreement from 
weak to strong coupling. The FCB suppression of $\bar{I}$ at large $\l$ 
as well as the phonon-assisted current steps at 
$V=2\w_{0}$ are correctly reproduced. Turning on the ee interaction 
the Coulomb deblocking takes place and $\bar{I}$ increases for all $V$. 
We still observe phonon-assisted steps but, unexpectedly, they bend 
{\em downward} giving rise to regions of  NDC. 
This phenomenon is our second main finding and is driven by the 
competition between  ee and ep 
interactions (no NDC for $U=0$ or $\l=0$).
By further increasing the bias a crossover occurs: the steps 
are attenuated and the current acquires a power-law decay 
$\bar{I}\sim V^{\b-1}$. In this region the system behaves as if the 
ep coupling were zero~\cite{boulat,irlm}.

{\it NDC.---}
We investigate further the NDC aspect by calculating the $dI/dV$ 
as a function of voltage $V$ and gate $\tilde{\e}_{d}$. NDC regions 
have been observed in QDs formed  between 
the defects of a carbon nanotube (CNT)~\cite{vonoppen}. Even though 
theoretical studies have so far been focussed on the ep 
coupling~\cite{vonoppen,sassetti}, the
left/right portion of the CNT screens the charge 
accumulated on 
the QD. Our Hamiltonian represents the simplest generalization of 
previous models to include this screening effect.
We use parameters from the literature: $\w_{0}=1$ meV, $\l=1.82$ 
meV, $a=2.46$ \AA,
$v_{F}=8.1\times10^{5}$ m/s, $\G=0.1$ meV, and
$\L=0.1$ eV~\cite{cntu}. 
For the ee coupling we take $U<5$ eV, since in CNTs
the on-site repulsion is $\sim 5$ eV~\cite{auger}.
In Fig. \ref{fig4} we show the contour plot of the
$dI/dV$ for three different $U$s. The $U=0$ case 
accurately reproduces the FCB diamonds obtained within the 
rate equations approach~\cite{fcb} and later observed in 
experiments~\cite{vonoppen}. 
However, no signatures of NDC are found. 
For $U=1$ eV, instead, spots of NDC 
appear inside the diamonds, in qualitative 
agreeement with the experiment. Increasing $U$ even further 
the NDC regions expand, and  
horizontal stripes of large conductance emerge. 
However, these stripes should be suppressed by the strong, local 
repulsion (not considered here) responsible for the standard Coulomb 
blockade.

{\it Conclusions.---}
We derived an approximate, yet accurate, formula for the electrical 
current through a QD with ep and ee coupling. Screening and polaronic features are 
transparently incorporated, rendering the physical interpretation 
direct and intuitive.
The competition between FCB and Coulomb deblocking leads to the novel 
effect of NDC regions in the $dI/dV$. This mechanism occurs in QD 
weakly coupled to low-dimensional leads, like those recently realized 
with CNT.

We acknowledges funding by MIUR FIRB 
grant No. RBFR12SW0J.



\begin{thebibliography}{99}
\bibitem{nota4}
We briefly discuss the differences with an alternative 
approach to deal with
ep interactions ($U=0$)~\cite{galperin}. Our
correlator between the $f$ fields is not dressed by 
virtual tunneling processes whereas in Ref. \cite{galperin} this 
dressing is accounted for up to second-order in a self-consistent 
manner. However, in Ref. \cite{galperin} the Green's function 
satisfies Eq. (\ref{eom4}) in which the $G$ under the integral sign 
is replaced by the noninteracting $G_{0}$. Consequently the 
noninteracting case ($\l=0$) is not recovered.
\bibitem{nota2}
The retarded component of $\Sigma_{\a}$ in Eq. (\ref{meir}) is extracted 
from 
$\Sigma_{\a}^{\rm R}(\w)=\int \frac{d \w'}{2\pi i} 
\frac{\Sigma_{\a}^{<}(\w')-\Sigma_{\a}^{>}(\w')}{\w-\w'+i\eta}$.
\end{thebibliography}
\end{document}